\newif\ifplainstyle
\plainstyletrue
\plainstylefalse

\newif\ifjhepstyle
\jhepstyletrue

\newif\ifprstyle
\prstyletrue
\prstylefalse

\ifprstyle
	\documentclass[twocolumn,nofootinbib]{revtex4-1}
\else
	\documentclass[11pt,a4paper]{article}
\fi

\ifjhepstyle
	\usepackage{jheppub}
	\usepackage{amsfonts}
	\usepackage{verbatim}
	\usepackage{float}
	\usepackage{color}
	
	\usepackage{array}
	\newcolumntype{C}[1]{>{\centering\arraybackslash$}p{#1}<{$}}
	\makeatletter
	\def\@fpheader{\phantom{Prepared for submission to JHEP}}
	\makeatother
\else	
 	\ifprstyle
		\usepackage{verbatim}
		\usepackage{amsmath,amsfonts,amssymb}
		\usepackage[colorlinks=true
                	,urlcolor=blue
                	,anchorcolor=blue
                	,citecolor=blue
                	,filecolor=blue
                	,linkcolor=blue
                	,menucolor=blue
                	]{hyperref}
	\else
            	\usepackage{verbatim}
            	\usepackage{cite}
            	\usepackage{setspace}
            	\usepackage[top=2.5cm, bottom=2.75cm, left=2.5cm, right=2.5cm]{geometry}
            	\usepackage{amsmath,amsfonts,amssymb}
            	\usepackage[colorlinks=true
            	,urlcolor=blue
            	,anchorcolor=blue
            	,citecolor=blue
            	,filecolor=blue
            	,linkcolor=blue
            	,menucolor=blue
            	,linktoc=page
            	]{hyperref}
            	\usepackage{float}
            	\restylefloat{table}
            	
            	\numberwithin{equation}{section}
	\fi
\fi

\allowdisplaybreaks

\usepackage{empheq}
\usepackage{tikz}
\usepackage{mathtools}
\usepackage{pgfplots}
\usepackage{subcaption}
\usepackage{bbold}
\usepackage{transparent}
\usepackage{bm}
\usepackage{enumitem}
\usepackage[normalem]{ulem}

\makeatletter
\let\save@mathaccent\mathaccent
\newcommand*\if@single[3]{%
  \setbox0\hbox{${\mathaccent"0362{#1}}^H$}%
  \setbox2\hbox{${\mathaccent"0362{\kern0pt#1}}^H$}%
  \ifdim\ht0=\ht2 #3\else #2\fi
  }
\newcommand*\rel@kern[1]{\kern#1\dimexpr\macc@kerna}
\newcommand*\widebar[1]{\@ifnextchar^{{\wide@bar{#1}{0}}}{\wide@bar{#1}{1}}}
\newcommand*\wide@bar[2]{\if@single{#1}{\wide@bar@{#1}{#2}{1}}{\wide@bar@{#1}{#2}{2}}}
\newcommand*\wide@bar@[3]{%
  \begingroup
  \def\mathaccent##1##2{%
    \let\mathaccent\save@mathaccent
    \if#32 \let\macc@nucleus\first@char \fi
    \setbox\z@\hbox{$\macc@style{\macc@nucleus}_{}$}%
    \setbox\tw@\hbox{$\macc@style{\macc@nucleus}{}_{}$}%
    \dimen@\wd\tw@
    \advance\dimen@-\wd\z@
    \divide\dimen@ 3
    \@tempdima\wd\tw@
    \advance\@tempdima-\scriptspace
    \divide\@tempdima 10
    \advance\dimen@-\@tempdima
    \ifdim\dimen@>\z@ \dimen@0pt\fi
    \rel@kern{0.6}\kern-\dimen@
    \if#31
      \overline{\rel@kern{-0.6}\kern\dimen@\macc@nucleus\rel@kern{0.4}\kern\dimen@}%
      \advance\dimen@0.4\dimexpr\macc@kerna
      \let\final@kern#2%
      \ifdim\dimen@<\z@ \let\final@kern1\fi
      \if\final@kern1 \kern-\dimen@\fi
    \else
      \overline{\rel@kern{-0.6}\kern\dimen@#1}%
    \fi
  }%
  \macc@depth\@ne
  \let\math@bgroup\@empty \let\math@egroup\macc@set@skewchar
  \mathsurround\z@ \frozen@everymath{\mathgroup\macc@group\relax}%
  \macc@set@skewchar\relax
  \let\mathaccentV\macc@nested@a
  \if#31
    \macc@nested@a\relax111{#1}%
  \else
    \def\gobble@till@marker##1\endmarker{}%
    \futurelet\first@char\gobble@till@marker#1\endmarker
    \ifcat\noexpand\first@char A\else
      \def\first@char{}%
    \fi
    \macc@nested@a\relax111{\first@char}%
  \fi
  \endgroup
}
\makeatother

\newcommand{\ThisIsTheTitle}{Strings on warped AdS$_3$ via $\bm{T\bar{J}}$ deformations } 
\newcommand{\ThisIsAuthorOne}{Luis Apolo}
\newcommand{\ThisIsEmailOne}{apolo@math.tsinghua.edu.cn}
\newcommand{\ThisIsAuthorTwo}{and Wei Song}
\newcommand{\ThisIsEmailTwo}{wsong@math.tsinghua.edu.cn}
\newcommand{\ThisIsTheAffiliation}{Yau Mathematical Sciences Center, Tsinghua University, Beijing 100084, China}
\newcommand{\TheseAreTheKeywords}{}

\newcommand{\ThisIsTheAbstract}{We study a toy model of the Kerr/CFT correspondence using string theory on AdS$_3 \times S^3$. We propose a single trace irrelevant deformation of the dual CFT generated by a vertex operator with spacetime dimensions $(2,1)$. This operator shares the same quantum numbers as the integrable $T\bar{J}$ deformation of two-dimensional CFTs where $\bar{J}$ is a chiral $U(1)$ current. We show that the deformation is marginal on the worldsheet and that the target spacetime is deformed to null warped AdS$_3$ upon dimensional reduction. We also calculate the spectrum of the deformed theory on the cylinder and compare it to the field theory analysis of $T\bar{J}$-deformed CFTs.}

\ifjhepstyle
\title{\ThisIsTheTitle}

\author{\ThisIsAuthorOne}
\author{\ThisIsAuthorTwo}

\affiliation{\ThisIsTheAffiliation}

\emailAdd{\ThisIsEmailOne}
\emailAdd{\ThisIsEmailTwo}

\abstract{\ThisIsTheAbstract} 

\keywords{\TheseAreTheKeywords}
\fi

\begin{document}


\ifjhepstyle
\maketitle
\flushbottom
\fi

\long\def\symfootnote[#1]#2{\begingroup%
\def\thefootnote{\fnsymbol{footnote}}\footnote[#1]{#2}\endgroup}

\def\rednote#1{{\color{red} #1}}
\def\bluenote#1{{\color{blue} #1}}

\def\({\left (}
\def\){\right )}
\def\lb{\left [}
\def\rb{\right ]}
\def\lB{\left \{}
\def\rB{\right \}}

\def\Int#1#2{\int \textrm{d}^{#1} x \sqrt{|#2|}}
\def\Bra#1{\left\langle#1\right|} 
\def\Ket#1{\left|#1\right\rangle}
\def\BraKet#1#2{\left\langle#1|#2\right\rangle} 
\def\Vev#1{\left\langle#1\right\rangle}
\def\Vevm#1{\left\langle \Phi |#1| \Phi \right\rangle}\def\bbox{\bar{\Box}}
\def\til#1{\tilde{#1}}
\def\wtil#1{\widetilde{#1}}
\def\ph#1{\phantom{#1}}

\def\ra{\rightarrow}
\def\la{\leftarrow}
\def\lra{\leftrightarrow}
\def\p{\partial}
\def\diff{\mathrm{d}}

\def\sinh{\mathrm{sinh}}
\def\cosh{\mathrm{cosh}}
\def\tanh{\mathrm{tanh}}
\def\coth{\mathrm{coth}}
\def\sech{\mathrm{sech}}
\def\csch{\mathrm{csch}}

\def\a{\alpha}
\def\b{\beta}
\def\g{\gamma}
\def\d{\delta}
\def\e{\epsilon}
\def\ve{\varepsilon}
\def\k{\kappa}
\def\l{\lambda}
\def\n{\nabla}
\def\om{\omega}
\def\s{\sigma}
\def\t{\theta}
\def\z{\zeta}
\def\vp{\varphi}

\def\ss{\Sigma}
\def\dd{\Delta}
\def\GG{\Gamma}
\def\ll{\Lambda}
\def\tt{\Theta}

\def\A{{\cal A}}
\def\B{{\cal B}}
\def\C{{\cal C}}
\def\cE{{\cal E}}
\def\D{{\cal D}}
\def\F{{\cal F}}
\def\H{{\cal H}}
\def\I{{\cal I}}
\def\J{{\cal J}}
\def\K{{\cal K}}
\def\L{{\cal L}}
\def\N{{\cal N}}
\def\O{{\cal O}}
\def\P{{\cal P}}
\def\cS{{\cal S}}
\def\W{{\cal W}}
\def\X{{\cal X}}
\def\Z{{\cal Z}}

\def\mfa{\mathfrak{a}}
\def\mfb{\mathfrak{b}}
\def\mfc{\mathfrak{c}}
\def\mfd{\mathfrak{d}}

\def\we{\wedge}
\def\re{\textrm{Re}}

\def\tilw{\tilde{w}}
\def\tile{\tilde{e}}

\def\tilL{\tilde{L}}
\def\tilJ{\tilde{J}}

\def\zz{\bar z}
\def\xx{\bar x}
\def\yy{\bar y}
\def\xp{x^{+}}
\def\xm{x^{-}}

\def\bp{\bar{\p}}
\def\wei#1{{\color{red}#1}}

\def\VirU1{Vir \times U(1)}
\def\VirSL2R{\mathrm{Vir}\otimes\widehat{\mathrm{SL}}(2,\mathbb{R})}
\def\U1{U(1)}
\def\u1{U(1)}
\def\SL2R{\widehat{\mathrm{SL}}(2,\mathbb{R})}
\def\sl2r{\mathrm{SL}(2,\mathbb{R})}
\def\by{\mathrm{BY}}

\def\RR{\mathbb{R}}

\def\tr{\mathrm{Tr}}
\def\bnabla{\overline{\nabla}}

\def\sint{\int_{\ss}}
\def\dsint{\int_{\p\ss}}
\def\hint{\int_{H}}

\newcommand{\eq}[1]{\begin{align}#1\end{align}}
\newcommand{\eqst}[1]{\begin{align*}#1\end{align*}}
\newcommand{\eqsp}[1]{\begin{equation}\begin{split}#1\end{split}\end{equation}}

\newcommand{\absq}[1]{{\textstyle\sqrt{\left |#1\right |}}}



\ifprstyle
\title{\ThisIsTheTitle}

\author{\ThisIsAuthorOne}
\email{\ThisIsEmailOne}

\author{\ThisIsAuthorTwo}
\email{\ThisIsEmailTwo}

\affiliation{\ThisIsTheAffiliation}


\begin{abstract}
\ThisIsTheAbstract
\end{abstract}


\maketitle

\fi

\ifplainstyle
\begin{titlepage}
\begin{center}

\ph{.}

\vskip 4 cm

{\Large \bf \ThisIsTheTitle}

\vskip 1 cm

\renewcommand*{\thefootnote}{\fnsymbol{footnote}}

{{\ThisIsAuthorOne}\footnote{\ThisIsEmailOne} } and {{\ThisIsAuthorTwo}\footnote{\ThisIsEmailTwo}}

\renewcommand*{\thefootnote}{\arabic{footnote}}

\setcounter{footnote}{0}

\vskip .75 cm

{\em \ThisIsTheAffiliation}

\end{center}

\vskip 1.25 cm

\begin{abstract}
\noindent \ThisIsTheAbstract
\end{abstract}

\end{titlepage}

\newpage

\fi

\ifplainstyle
\tableofcontents
\noindent\hrulefill
\bigskip
\fi

\section{Introduction} \label{se:intro}
Black holes play an essential role in understanding quantum gravity. In string theory, Strominger and Vafa~\cite{Strominger:1996sh} provided the first microscopic interpretation of the Bekenstein-Hawking entropy for BPS black holes, which later developed into the holographic duality between string theory on AdS$_3\times S^3\times \mathcal{M}^4$ and two-dimensional conformal field theories (CFTs).  A similar attempt towards understanding real world black holes was initiated with the Kerr/CFT correspondence~\cite{Guica:2008mu}. Kerr/CFT conjectured the existence of a holographic dual to (near) extremal black holes, the latter of which have been reported to exist in nature~\cite{McClintock:2006xd}. The near-horizon region of four-dimensional extremal Kerr black holes (NHEK) features a $U(1)_L \times \widebar{SL(2,R)}_R$ isometry group~\cite{Bardeen:1999px}, a property shared by a three dimensional geometry dubbed warped AdS$_3$ (WAdS$_3$)~\cite{Anninos:2008fx}. WAdS$_3$ appears in several contexts: as a section of NHEK at fixed polar angle, as a factor in the near horizon geometry of six-dimensional black strings and five-dimensional black holes~\cite{Dias:2007nj, Bredberg:2009pv, Guica:2010ej,Bouchareb:2013dka,Bouchareb:2014fxa}, and as a solution to some three dimensional theories of gravity~\cite{Vuorio:1985ta,Percacci:1986ja,Moussa:2003fc, Bouchareb:2007yx, Anninos:2008fx, Moussa:2008sj, Clement:2009gq, Detournay:2012dz}. Holographic dualities for WAdS$_3$ spacetimes provide simpler toy models for the Kerr/CFT correspondence. Furthermore, since neither NHEK nor WAdS$_3$ are asymptotically AdS, the proposed dualities explore holography beyond AdS spacetimes. 

Without resorting to string theory, the Kerr/CFT correspondence and related proposals for WAdS$_3$ have been successful in reproducing universal properties of black holes such as the Bekenstein-Hawking entropy and greybody factors~\cite{Guica:2008mu,Bredberg:2009pv,Detournay:2012pc,Castro:2010fd,Song:2017czq}.  Despite its name, however, there is strong evidence that the dual field theory in Kerr/CFT is not a local QFT~\cite{Guica:2010sw}. To gain a better understanding of these theories it is useful to embed the Kerr/CFT correspondence in string theory~\cite{Guica:2010ej, Compere:2010uk}. In this context, Kerr/CFT can be understood as an irrelevant deformation induced by a $(h,\bar{h})=(2,1)$ operator~\cite{Guica:2010ej, Compere:2010uk, ElShowk:2011cm,Song:2011sr}. Relatedly, a geometry containing a WAdS$_3$ factor can be obtained from string theory on AdS$_3$ via marginal deformations of the worldsheet~\cite{Israel:2004vv,Detournay:2005fz,Detournay:2010rh}, or a sequence of solution-generating transformations~\cite{ElShowk:2011cm, Song:2011sr, Bena:2012wc}. For example, a TsT transformation~\cite{Lunin:2005jy,Maldacena:2008wh} mixing $U(1)$ currents from the AdS$_3$ and an $S^3$ factor can be interpreted as a marginal worldsheet deformation that leads to a WAdS$_3$ geometry~\cite{Azeyanagi:2012zd}. We will see that this transformation corresponds to an irrelevant (2,1) deformation of the dual CFT.

While generally difficult to deal with, recently a class of irrelevant and Lorentz invariant deformations has been shown to be integrable~\cite{Smirnov:2016lqw,Cavaglia:2016oda}. In particular, the so-called $T{\bar T}$ deformation of a two-dimensional CFT interpolates between an IR fixed point and a nonlocal QFT in the UV. The holographic description of the deformed theory depends on the single or double trace character of the deformation, as well as on the sign of its dimensionful coupling $\mu$. For example, in~\cite{McGough:2016lol} a double trace deformation with $\mu< 0$ was argued to be dual to AdS$_3$ with a finite cutoff at radius $r_c \propto 1/\sqrt{|\mu|}$. In contrast, refs.~\cite{Giveon:2017nie,Giveon:2017myj} considered a single trace deformation which, unlike its double trace analog, changes the background metric. The single trace $T\bar{T}$ deformation corresponds to a marginal deformation of string theory that interpolates between AdS$_3$ and linear dilaton backgrounds~\cite{Israel:2003ry}. The latter are vacua of Little String Theory which features a Hagedorn spectrum that matches the spectrum found in the deformed theory~\cite{Giveon:2017nie,Giveon:2017myj}. For related work on $T\bar{T}$ deformations see e.g.~\cite{Dubovsky:2017cnj,Shyam:2017znq,Asrat:2017tzd,Giribet:2017imm,Kraus:2018xrn,Cottrell:2018skz,Chakraborty:2018kpr,Dubovsky:2018bmo,Aharony:2018vux,Baggio:2018gct,Dei:2018mfl}.

Integrable deformations of two-dimensional CFTs are not restricted to Lorentz invariant operators. In particular, the irrelevant deformations relevant to holography for NHEK and WAdS$_3$ spacetimes must necessarily break Lorentz invariance. The simplest example of such deformations is the (2,1) $T\bar{J}$ deformation proposed in~\cite{Guica:2017lia} where $\bar{J}$ is a $\widebar{U(1)}_R$ current.\footnote{Here we study (2,1) instead of the (1,2) deformations originally considered in~\cite{Guica:2017lia,Bzowski:2018pcy}. This convention is chosen to maximize the overlap with previous work on string theory on warped AdS$_3 \times S^3$ backgrounds~\cite{Azeyanagi:2012zd}.}  The $T\bar{J}$ deformation is solvable and breaks the global $SL(2,R)_L \times \widebar{SL(2,R)}_R$ symmetries of the CFT down to $U(1)_L\times \widebar{SL(2,R)}_R$. The resulting theory is nonlocal, a feature shared by the holographic duals to NHEK and WAdS$_3$ beyond the matching of global symmetries. In analogy to the $T\bar{T}$ deformation, double trace $T\bar{J}$-deformed CFTs were shown to be dual to AdS$_3$~\cite{Bzowski:2018pcy} with modified boundary conditions similar to~\cite{Compere:2013bya}. However, in order to get a toy model of Kerr/CFT it is also necessary to deform the bulk metric to WAdS$_3$. To obtain WAdS$_3$ we therefore consider a single trace $T\bar{J}$ deformation along the lines of~\cite{Giveon:2017nie,Giveon:2017myj}.
 
In this paper, we study a toy model of Kerr/CFT in string theory via an irrelevant (2,1) deformation. Our starting point is string theory on AdS$_3\times S^3\times {\cal M}^4$  backgrounds supported by NS-NS flux. The theory features (2,1) vertex operators that correspond to observables in the dual two-dimensional CFT. These operators are given by~\cite{Kutasov:1999xu}
 \eq{
  A^a (x,\bar{x}) = \frac{1}{2} \int d^2 z \Big [ \p_{{x}} {J}({x};{z}) \p_{{x}} +2 \p_{{x}}^2 {J}({x};{z}) \Big ] \Phi_1(x,\bar{x};z,\bar{z}) \bar{k}^a(\bar{z}), \label{ktintro}
  }
where $(x,\bar x)$ are the coordinates of the dual CFT, $(z,\bar{z})$ are the coordinates on the worldsheet, $J(x;z)$ is an $x$-dependent linear combination of $SL(2,R)_L$ worldsheet currents while $\bar{k}^{a}(\bar{z})$ is an $\widebar{SU(2)}_R$ worldsheet current, and $\Phi_1(x,\bar{x}; z,\bar{z})$ is an $SL(2,R)_L \times \widebar{SL(2,R)}_R$ primary. 

\vspace{5pt}

\noindent{\bf In the dual CFT}, we conjecture that the (2,1) vertex operator \eqref{ktintro} corresponds to a single trace $T\bar{K}$ deformation where $\bar{K}$ is one of the $SU(2)$ currents in the dual field theory. As evidence, we show that the single trace deformation satisfies the appropriate OPEs with other conserved currents of the dual CFT. These OPEs are similar to those satisfied by the double trace deformation, except for terms proportional to the central extensions. We will see that the $T\bar{K}$ deformation reduces to a marginal deformation of string theory bilinear in worldsheet currents~\cite{Azeyanagi:2012zd}. Furthermore, by placing the deformed theory on the cylinder we derive its finite-size spectrum. Denoting by $\l$ the dimensionful deformation parameter, $E_L(\l)$ and $E_R(\l)$  the left and right-moving energies, $P(\lambda)$ the deformed momentum, and $\bar{Q}(\l)$ the expectation value of the $\widebar{U(1)}_R$ current, we find
  \eq{
    2P(\l) &\equiv E_L(\l) - E_R(\l) = E_L(0) - E_R(0) = 2 P(0), \\
    E_L(\l) & =\dfrac{\l \bar{q} +k| w |R}{2 \l^2} - \dfrac{1}{2 \l^2} \sqrt{\vphantom{\tfrac{1}{2}} (\l \bar{q}+k |w| R)^2 - 4 k |w| R \,\l^2 E_L(0)\,}, \qquad w < 0, \label{leftenergyintro} \\
  \bar{Q}(\l) & = {\bar{q}}{} - 2\l E_L(\l).
  }
In eq.~\eqref{leftenergyintro} $w$ is the spectral flow parameter, i.e.~the string winding number, $\bar{q}$ is the charge of the undeformed $\widebar{U(1)}_R$ symmetry, and $k = \ell_{AdS}^2/ \ell_{s}^2$ where $\ell_{AdS}$ and $\ell_s$ are the AdS and string scales, respectively.

\vspace{5pt}

\noindent{\bf On the gravity side}, we show that the deformed theory corresponds to a string theory propagating on the warped AdS$_3\times S^3$ background originally found in~\cite{Azeyanagi:2012zd}. The deformation breaks the $SL(2,R)_L \times\widebar{SL(2,R)}_R \times SU(2)_L \times \widebar{SU(2)}_R$ isometry of AdS$_3\times S^3$ down to $U(1)_L \times \widebar{SL(2,R)}_R$ and $SU(2)_L \times \widebar{U(1)}_R$. Furthermore, the warped AdS$_3\times S^3$ geometry reduces to null warped AdS$_3$ after dimensional reduction. Thus, the single trace $T\bar{K}$ deformation reproduces the kind of spacetimes and symmetries featured in the near horizon limit of extremal black holes.

The paper is organized as follows. In Section~\ref{se:strings} we review aspects of string theory on AdS$_3$. In particular we review the construction of vertex operators corresponding to the conserved currents in the dual CFT. The conjectured single trace $T\bar{K}$ deformation is considered in Section~\ref{se:warped}. Therein we show that the deformation is marginal on the worldsheet and that it describes strings on a warped AdS$_3$ spacetime. Finally, in Section~\ref{se:spectrum} we derive the finite-size spectrum of the dual CFT via spectral flow.

\bigskip

\noindent {\bf{Note added:}} while this work was in progress we learned of ref.~\cite{Chakraborty:2018vja} which has some overlap with our work. 


\section{String theory on AdS$_3$} \label{se:strings}

In this section we review some of the basic ingredients of string theory on AdS$_3$ that are necessary to construct the single trace $T\bar{K}$ deformation. We begin by considering the currents of the worldsheet theory and then use these to write down the currents in the dual conformal field theory. 


\subsection{Action and worldsheet currents}

Let us consider string theory on the following background
  \eq{
  AdS_3 \times S^3 \times {\cal M}^4, \label{background}
  }
supported by NS-NS flux. The compact, four-dimensional manifold ${\cal M}^4$ in eq.~\eqref{background} does not play a role in our story and is henceforth omitted from the discussion. The background metric and $B$ fields are respectively given by 
  \eq{
  ds^2 &= k \Big \{ d \phi^2 + e^{2\phi} d\g d\bar{\g}  + \frac{1}{4} \Big [ d\t^2 + \sin^2 \t \, d \vp^2 + \( d \psi + \cos\t \, d \vp \)^2 \Big ] \Big \},  \label{metric} \\
 B &= \frac{k}{4} \Big \{2 e^{2\phi}  d \bar{\g} \we d\g  + \cos \t \, d \psi \we d \vp \Big \}, \label{bfield}
  }
where $\(\phi, \g=\gamma^1+\gamma^0, \bar{\g}=\gamma^1-\gamma^0 \)$ denote the Poincar\'e coordinates of AdS$_3$, $\(\psi, \vp, \t\)$ parametrize the $S^3$ as a Hopf fibration, and the ratio of the AdS ($\ell_{AdS}$) and string ($\ell_s$) scales is assumed to be large $k = \ell_{AdS}^2/\ell_s^2 \gg 1$. The bosonic part of the worldsheet action is thus given by
  \eq{
  S = k \int d^2 z \bigg \{ \p \phi \bar{\p} \phi + e^{2\phi} \bar{\p} \g \p \bar{\g} + \frac{1}{4}  \Big [ \p \t \bp \t + \p \vp \bp \vp + \( \bp \psi + 2 \cos \t \, \bp \vp  \) \p \psi \Big ]  \bigg \}, \label{action}
  }
where we have set the string length to $\ell_s = 1$, and $(\p$,  $\bar{\p})$ denote derivatives with respect to the worldsheet coordinates $(z,\bar{z})$.



The action~\eqref{action} features both left and right-moving $SL(2,R) \times SU(2)$ Kac-Moody symmetries. Following the conventions of~\cite{Azeyanagi:2012zd} we denote quantities in the right-moving sector with a bar. The left-moving $SL(2,R)_L$ worldsheet currents are given by
 \eq{
\!\! j^-(z) \!=\! - k \, e^{2\phi} \p \bar{\g}, \quad j^3(z) \! = \!-k \big ( e^{2\phi} \g \p \bar{\g} - \p \phi \big), \quad j^+(z) \! =\! -k \big ( e^{2\phi} \g^2 \p \bar{\g} - 2\g \p \phi - \p \g \big ),\label{sl2rcurrents}
 }
and satisfy the OPEs
  \eq{
  j^+(z)j^-(w) &\sim \frac{k}{(z-w)^2} + \frac{2 j^3(w)}{z-w}, \label{jope1} \\
  j^3(z) j^3(w) &\sim -\frac{k/2}{(z-w)^2}, \\
  j^3(z) j^{\pm}(w) &\sim  \pm \frac{j^{\pm}(w)}{z-w}, \label{jope3}
  }
where $\sim$ denotes equality up to regular terms. On the other hand the right-moving $\widebar{SU(2)}_R$ currents read
  \eq{
  \bar{k}^3(\bar{z}) = -i \frac{k}{2} \big ( \bp \psi  + \cos\t \,\bp \vp \big), \qquad \bar{k}^{\pm}(\bar{z}) = \frac{k}{2} e^{\mp i \psi} \big ( \bp \t \pm i \sin\t \, \bp\vp \big ). \label{su2currents}
  }
Their OPEs are given by
  \eq{
  \bar{k}^a (\bar{z}) \bar{k}^b(\bar{w}) \sim \frac{\tfrac{k}{2} \eta^{ab}}{(\bar{z}-\bar{w})^2} + \frac{i f^{abc} \eta_{cd} \bar{k}^d(\bar{w})}{\bar{z}-\bar{w}}, \label{kope}
  }
where $\eta_{ab}$ and $f^{abc}$ are respectively the Cartan-Killing metric and structure constant of $SU(2)$. 

Similarly, the currents for the $\widebar{SL(2,R)}_R$ and $SU(2)_L$ symmetries, namely $\bar{j}^i(\bar{z})$ and $k^a(z)$, are obtained from eqs.~\eqref{sl2rcurrents} and~\eqref{su2currents} by the following substitutions
  \eq{
  \qquad \p \lra \bar{\p}, \qquad \g \lra \bar{\g}, \qquad \psi \lra \vp.
  }
  %


\subsection{Spacetime currents} 
 
We now turn to the conserved currents of the CFT dual to string theory on AdS$_3$, specifically to the vertex operators for the stress energy tensor and $\widebar{SU(2)}_R$ currents. Following~\cite{Kutasov:1999xu} we refer to these currents as ``spacetime''  currents, in contrast to the worldsheet currents considered in the previous section. The spacetime currents depend on auxiliary variables $(x, \bar{x})$ that are interpreted as living on the boundary of AdS$_3$, i.e.~these are the coordinates of the dual CFT. First, we introduce the following parametrization of the left-moving $SL(2,R)_L$ currents~\cite{Kutasov:1999xu}
  \eq{
  J(x;z) = 2x j^3(z) - j^+(z) -x^2 j^-(z), \label{spacetimesl2}
  }
whose OPE follows from eqs.~\eqref{jope1} --~\eqref{jope3} and is given by
  \eq{
  J(x;z) J(y;w) \sim k\frac{(x-y)^2}{(z-w)^2} + \frac{1}{z-w} \big [ (x-y)^2 \p_y + 2 (x - y) \big ] J(y; w). \label{JJope}
  }
The right moving operator $\bar{J}(\bar{x};\bar{z})$ is obtained by letting $j^i (z) \ra \bar{j}^i(\bar{z})$ and $x \ra \bar{x}$ in eq.~\eqref{spacetimesl2}. Crucially, $J(x;z)$ carries worldsheet dimension $(1,0)$ and spacetime dimension $(-1,0)$. The latter follows from the fact that $-\oint dz j^3(z)$, $-\oint dzj^+(z)$, and $-\oint dzj^-(z)$ correspond to the $L_0$, $L_{+1}$, and $L_{-1}$ modes of the spacetime $SL(2,R)_L$ symmetry. 

A fundamental ingredient in the construction of vertex operators is the field $\Phi_h(x,\bar{x};z,\bar{z})$ defined by~\cite{Kutasov:1999xu}
  \eq{
  \Phi_h(x,\bar{x};z,\bar{z}) = \frac{1}{\pi} \bigg [ \frac{1}{ (\g-x) (\bar{\g} - \bar{x}) e^{\phi} + e^{-\phi}} \bigg ]^{2 h}. \label{phi}
  }
This operator is a primary under the  $SL(2,R)_L \times \widebar{SL(2,R)}_R$ symmetries of both the worldsheet and spacetime CFTs. $\Phi_h(x,\bar{x};z,\bar{z})$ may be interpreted as a bulk-to-boundary propagator whose worldsheet and spacetime dimensions are respectively given by $(\dd,\dd)$ and $(h,h)$ with $\dd = - h(h-1)/(k-2)$. Its OPE with the $SL(2,R)_L$ current $J(x;z)$ reads
  \eq{
  J(x;z) \Phi_h(y,\bar{y};w,\bar{w}) \sim \frac{1}{z-w} \big [ (x-y)^2 \p_y - 2 (x - y) \big ] \Phi_h(y,\bar{y};w,\bar{w}), \label{Jphiope}
  }
with a similar expression in terms of right-moving coordinates valid for its OPE with $\bar{J}(\bar{x};\bar{z})$.

Finally, the left-moving component of the stress energy tensor in the dual CFT is given, up to BRST exact terms, by 
  \eq{
  T(x) = -\frac{1}{2k} \int d^2z \Big [ \p_x J(x;z) \p_x + 2 \p_x^2 J(x;z) \Big ] \Phi_1(x,\bar{x};z,\bar{z}) \bar{J}(\bar{x};\bar{z}). \label{stresstensor}
  }
As expected, this is a vertex operator with spacetime dimension $(2,0)$. Similarly, the spacetime $\widebar{SU(2)}_R$ currents are given in terms of the worldsheet currents ${\bar k}^a({\bar z})$ by
  \eq{
  {\bar K}^a(\bar{x}) = -\frac{1}{k} \int d^2z {\bar k}^a(\bar z)\Phi_1(x,\bar{x};z,\bar{z}) J(x,z). \label{spacetimesu2}
  } 
It is not difficult to check that ${\bar K}^a(\bar{x})$ is an operator of dimension $(0,1)$. The spacetime currents discussed above are conserved within correlation functions and satisfy OPEs that lead to spacetime left and right-moving Virasoro-${SU(2)}$-Kac-Moody algebras. In particular, we have
  \eq{
  T(x) T(y) &\sim \frac{3k I}{(x-y)^4} + \frac{2 T(y)}{(x-y)^2} +  \frac{\p_y T(y)}{x-y}, \label{spacetimeope1} \\
  \bar{T}(\xx) \bar{K}^a (\yy) &\sim \frac{\bar{K}^a(\yy)}{(\xx-\yy)^2} + \frac{\p_{\yy} \bar{K}^a(\yy)}{\xx-\yy}, \\
  \bar{K}^a(\xx) \bar{K}^b(\yy) &\sim \frac{\tfrac{k}{2} \eta^{ab} I}{(\xx-\yy)^2} + \frac{i f^{abc} \eta_{cd} \bar{K}^d(\yy)}{\xx-\yy}, \label{spacetimeope3}
  }
and similarly for the $\bar{T}(\bar{x})$ and $K^a(x)$ currents. The central charge and level in eqs.~\eqref{spacetimeope1} and~\eqref{spacetimeope3} are proportional to $I$, a multiple of the identity operator defined in~\cite{Kutasov:1999xu} by
  \eq{
  I  = \frac{1}{k^2} \int d^2z J(x;z) \Phi_1 (x, \bar{x}; z, \bar{z}) \bar{J}(\bar{x}; \bar{z}).\label{identity}
  }
Note that eq.~\eqref{identity} can take different values in different sectors of the theory~\cite{Giveon:2001up}.


\subsection{Holographic dual} 

String theory on $AdS_3\times S^3\times {\cal M}^4$ supported by NS-NS flux is conjectured to be dual to a deformation of the symmetric product orbifold~\cite{Argurio:2000tb}
  \eq{
  \big ( {\cal M}_{6k} \big )^p/S_p,  \label{symprod}
  }
where ${\cal M}_{6k}$ is a CFT with central charge $c = 6k$ that depends on the details of the internal manifold ${\cal M}^4$. In eq.~\eqref{symprod} $k$ denotes the level of the current algebra and $p$ counts the number of fundamental strings necessary to generate the background~\eqref{metric} (in contrast, $k$ counts the number of NS$_5$ branes).  
The total central charge of the dual CFT is $c_{\textrm{total}} = 6 k p$, in agreement with the worldsheet derivation of refs.~\cite{Giveon:1998ns,Kutasov:1999xu}. In particular, it has been argued that~\eqref{symprod} lies in the same moduli space as the D1D5 CFT. 

At critical values of $k$ --- $k = 3$ for the bosonic string, $k =1$ for the superstring --- there is evidence that the dual theory is at the orbifold point~\cite{Giribet:2018ada,Gaberdiel:2018rqv}. In contrast, for generic values of $k$ the holographic dual is expected to be a deformation of the symmetric product.



\section{Warped AdS$_3$ from $T\bar{K}$} \label{se:warped}

In this section we propose a single trace generalization of the $T\bar{K}$ deformation studied in~\cite{Guica:2017lia}. We find that this operator satisfies the appropriate OPEs with other conserved currents of the CFT dual to string theory on AdS$_3$. We show that the deformation is marginal on the worldsheet and that the target spacetime is deformed to null warped AdS$_3$ upon dimensional reduction.

\subsection{A single trace $T\bar{K}$ deformation}
 
As discussed in Section~\ref{se:intro}, deformations of two-dimensional CFTs by Lorentz-violating operators of weight $(2,1)$ play an important role in understanding the holographic duals to extremal Kerr black holes and warped AdS$_3$ spacetimes~\cite{Guica:2010sw, Compere:2010uk, ElShowk:2011cm, Song:2011sr}. Among the family of $(2,1)$ operators, deformations built from the a left-moving stress tensor and a right-moving $\widebar{U(1)}_R$ current appear to be the simplest. CFTs deformed by these operators have been shown to be integrable in~\cite{Guica:2017lia}. Their holographic duals were studied in~\cite{Bzowski:2018pcy} where the deforming operator was assumed to be double trace. In this picture, the bulk theory is still described by an asymptotically AdS$_3$ spacetime, but with modified boundary conditions similar to those of~\cite{Compere:2013bya}. In order to deform the bulk metric to WAdS$_3$ we need to find a single trace deformation. 

Let us consider possible $T\bar{K}$ deformations where $\bar{K}$ is a $\widebar{U(1)}_R$ current in the $\widebar{SU(2)}_R$ symmetry of the dual CFT. Clearly, this operator features spacetime dimension $(2,1)$. Besides breaking the Lorentz symmetry of the spacetime CFT, this deformation also breaks the affine $\widebar{SU(2)}_R$ symmetry down to an affine $\widebar{U(1)}_R$. There are two kinds of vertex operators one may study. One is a deformation by the double trace operator $T(x)\bar{K}^a(\bar{x})$. This is not a local deformation of the worldsheet CFT and similar in spirit to what has been already been discussed in~\cite{Bzowski:2018pcy}. 

To get a single trace deformation, we consider an alternative vertex operator of dimension $(2,1)$ that corresponds to a local deformation of the worldsheet theory. This operator can be built from the components making up the stress tensor and one of the $\widebar{SU(2)}_R$ currents~\cite{Kutasov:1999xu}. After a convenient normalization the single trace $T\bar{K}$ operator is given by
  \eq{
  A^a (x,\bar{x}) = \frac{1}{2} \int d^2 z \Big [ \p_{{x}} {J}({x};{z}) \p_{{x}} +2 \p_{{x}}^2 {J}({x};{z}) \Big ] \Phi_1(x,\bar{x};z,\bar{z}) \bar{k}^a(\bar{z}). \label{kt}
  }
  $A^a (x,\bar{x})$ differs from the left moving component of the stress tensor given in eq.~\eqref{stresstensor} on the last term, where the $\widebar{SL(2,R)}_R$ current $\bar{J}(\bar{x})$ has been replaced by the $\widebar{SU(2)}_R$ current $\bar{k}^a(\bar{z})$. Both currents have worldsheet dimension $(0,1)$ but feature different spacetime dimensions respectively given by  $(0,-1)$ and $(0,0)$. Following the lines of reasoning in~\cite{Giveon:2017nie}, we conjecture that eq.~\eqref{kt} corresponds to a single trace generalization of the $T\bar{K}$ deformation of~\cite{Guica:2017lia}. More precisely, if we deform the dual symmetric product CFT~\eqref{symprod}, 
the deformation is expected to be of the form of $\sum_i T_i(x) \bar{K}_i(\xx)$, where $T_i(x)$ and $\bar{K}_i(\xx)$ are currents on a single copy of ${\cal M}_{6k}$.

As evidence we show that, up to the central extensions, $A^a(x,\bar{x})$ shares the same form of the OPEs with the stress tensor and $\widebar{SU(2)}_R$ currents. Indeed, its OPE with the left-moving component of the stress tensor $T(x)$ is given by
  \eq{
  T(x) A^{a}(y,\bar{y}) \sim \frac{3k \bar{K}^{a}(\bar{y})}{(x - y)^4}  + \frac{2 A^a(y,\bar{y})}{(x-y)^2} + \frac{\p_y A^a(y,\bar{y})}{x - y}, \label{TAope}
  }
where we used the OPEs given in eqs.~\eqref{JJope} and~\eqref{Jphiope}, as well as~\cite{Kutasov:1999xu}
  \eq{
  \lim_{z\ra w} \Phi_1(x,\bar{x}; z,\bar{z}) \Phi_1(y,\bar{y}; w,\bar{w}) = \d^{(2)}(x-y) \Phi_1(y,\bar{y}; w,\bar{w}) + \O(z-w,\bar{z}-\bar{w}).
  }
Note that in eq~\eqref{TAope} we have ignored a term equal to $(x-y)^{-3} k \p_y\, \bar{K}^{a}(\bar{y})$ which is not identically zero but vanishes within correlation functions, consistent with the conservation of $\bar{K}^{a}(\bar{x})$. A similar term, namely $(x-y)^{-3}  k\, \p_y I$, appears in the $T(x) T(y)$ OPE where $I$ is a multiple of the identity operator~\eqref{identity} whose derivatives also vanish within correlation functions. One difference between the OPEs of $T(x)$ with the single and double trace $T\bar{K}$ deformations is found on their central extensions. The spacetime central charge in the $T(x) [T(y)\bar{K}^a(y)]$ OPE is given by $6k I$, cf.~eq.~\eqref{spacetimeope1}, where $I$ is proportional to the identity but takes different values on different sectors of the theory~\cite{Giveon:2001up}. In contrast, the central charge in the $T(x) A^a(x,\bar{x})$ OPE is simply $6k$.

There are two more OPEs we can compute which show that the single trace operator $A^a(x,\bar{x})$ behaves as desired. On the one hand we find that, up to contact terms, the OPE with the right moving component of the stress tensor $\bar{T}(\bar{x})$ is given by
  \eq{
  \bar{T}(\bar{x}) A^{a}(y,\bar{y}) \sim \frac{A^a(y,\bar{y})}{(\bar{x}-\bar{y})^2} + \frac{\p_{\bar{y}
  } A^a(y,\bar{y})}{\bar{x} - \bar{y}}, \label{TbarAope}
  }
as expected. On the other hand, using the OPEs of the worldsheet $\widebar{SU(2)}_R $ currents given in eq.~\eqref{kope} we obtain
  \eq{
  \bar{K}^a(\bar{x}) A^{b}(y,\bar{y}) \sim \frac{\tfrac{1}{2} k \eta^{ab} T(y)}{(\bar{x} - \bar{y})^2} + \frac{i f^{abc} \eta_{cd} A^{d}(y,\bar{y})}{\bar{x} - \bar{y}}, \label{KbarAope}
  }
where once again we have ignored all contact terms. The presence of the latter in the OPEs in eqs.~\eqref{TbarAope} and~\eqref{KbarAope} is due to the derivatives acting on $\Phi_1(x,\bar{x}; z, \bar{z})$ in the definition of $A^a(x,\bar{x})$. As before, we note that there is a difference in the central extension between the OPEs of $\bar{K}^a(\bar{x})$ with the single and double trace $T\bar{K}$ deformations. In the former the level is given by $kI$, cf.~eq.~\eqref{spacetimeope3}, while in the latter the level is instead given by $k$.
  
Thus, we expect the single trace generalization of the $T\bar{K}$ deformation on the dual CFT to be given by
  \eq{
  \int d^2x A^a(x,\bar{x}). \label{deformation}
  }
A crucial property of eq.~\eqref{deformation}, that is also shared by the single trace $T\bar{T}$ deformation considered in~\cite{Giveon:2017nie}, is that the $(x, \bar{x})$ dependence of the integrand is contained entirely in $\Phi_1(x,\bar{x};z,\bar{z})$. Indeed, using eq.~\eqref{spacetimesl2} we find, 
  \eq{
\int d^2x A^a(x,\bar{x}) = - \int d^2x \int d^2 z j^-(z) \bar{k}^a(\bar{z}) \Phi_1(x,\bar{x};z,\bar{z}), \label{kt2}
  }
where we have dropped total derivative terms which include potential short-distance singularities in the worldsheet coordinates originating from the $j^-(z) \Phi_1(x,\bar{x};z,\bar{z})$ OPE. Thus, since $\int d^2x \,\Phi_1(x,\bar{x};z,\bar{z}) = \frac{1}{\pi} \int d^2 x' (x' \bar{x}' + 1)^{-2}$ is a number, the single trace $T\bar{K}$ deformation corresponds to a \emph{marginal deformation} of the worldsheet theory that is given by
  \eq{
 \d S^{(a)} = \frac{2i\l}{k} \int d^2x A^a(x,\bar{x}) =  -\frac{2i \l}{k} \int d^2 z  {j}^{-}(z){\bar k}^a({\bar z}),\label{jtworldsheet}
  }
where we have absorbed $\int d^2 x \Phi_1(x,\bar{x};z,\bar{z})$ into $\lambda$. The coefficient of the deformation is chosen to guarantee that the deformed action considered in the next section is real and remains proportional to $k$, the latter of which is large in the semiclassical limit.


\subsection{Null warped AdS$_3$}

The deformation~\eqref{jtworldsheet} breaks the $SL(2,R)_L \times \widebar{SU(2)}_R$ symmetries of the action~\eqref{action}. Here we choose to align the $SU(2)$ index in~\eqref{jtworldsheet} along the $U(1)$ fibre of the 3-sphere, i.e.~we deform the worldsheet action by $A^3(x,\bar{x})$. Thus, the worldsheet action becomes
  \eq{
  S' & = S + \d S^{(3)} \\
  \begin{split}
  & = k \int d^2 z \bigg \{  \p \phi \bar{\p} \phi + e^{2\phi} \p \bar{\g} \Big [  \bar{\p} \g + \l (\bar{\p} \psi + \cos \t\,\bar{\p} \vp)  \Big ]  + \frac{1}{4} \( \bp \psi + 2 \cos \t \, \bp \vp  \) \p \psi \\
  & \phantom{= k \int d^2 z \bigg \{ }\,\,  +  \frac{1}{4} \( \p \t \bp \t + \p \vp \bp \vp \) \bigg \}. \label{deformedaction}
  \end{split}
  }
The deformed action~\eqref{deformedaction} corresponds to a string theory on the warped background
  \eq{
  \!\!ds^2 & = k \Big \{ d \phi^2 + e^{2\phi} d\bar{\g} \Big [ d\g  + \l (d \psi + \cos \t \,d \vp) \Big ] + \frac{1}{4} \Big [ d\t^2 + \sin^2 \t \, d \vp^2 + \( d \psi + \cos\t \,d \vp \)^2 \Big ]  \Big \}\label{deformedmetric}  \\
  B &= \frac{k}{4} \Big \{ 2 e^{2\phi} d\bar{\g} \we \Big[ d\g + \l (d \psi + \cos \t \,d \vp) \Big ] + \cos \t \, d \psi \we d \vp \Big \} .  \label{deformedbfield}
  }
The deformed metric~\eqref{deformedmetric} preserves only the $U(1)_L \times \widebar{SL(2,R)}_R \times SU(2)_L \times \widebar{U(1)}_R$
  subset of the original $SL(2,R)_L \times\widebar{SL(2,R)}_R \times SU(2)_L \times \widebar{SU(2)}_R$ isometries. The background described in eqs.~\eqref{deformedmetric} and~\eqref{deformedbfield} was originally obtained in ref.~\cite{Azeyanagi:2012zd} via the solution-generating TsT transformation of~\cite{Lunin:2005jy,Maldacena:2008wh} accompanied by an additional shift in the coordinates. More explicitly, starting from the AdS$_3 \times S^3$ background given in eqs.~\eqref{metric} and~\eqref{bfield}, one first T-dualizes along the $\g$ direction, shifts $\psi$ to $ \psi - \tfrac{2\l}{k} \g$, T-dualizes along $\g$ again and, crucially, shifts $\g$ to $\g + \tfrac{\l}{2} \psi$. We have thus shown that this generalized TsT transformation is equivalent to the $T\bar{K}$ deformation of the worldsheet action. 

Dimensional reduction on the 3-sphere yields a null warped AdS$_3$ background
  \eq{
  ds^2 = k \Big \{ d \phi^2 + e^{2\phi} d\g d\bar{\g} - \l^2 e^{4\phi} d \bar{\g} d \bar{\g} \Big \}.
  }
Thus, the single trace $T\bar{K}$ deformation of string theory on AdS$_3$ leads to one member of the family of metrics ubiquitous in the study of the near horizon limit of extremal black holes. The emergence of null warped AdS$_3$, as opposed to the timelike or spacelike solutions, follows from the fact that the original background was chosen to be AdS$_3$ in Poincar\'e coordinates. Indeed,  as explained in~\cite{Azeyanagi:2012zd}, a TsT transformation of a BTZ background yields spacelike warped AdS$_3$ black strings. This amounts to a deformation on the worldsheet by an operator of the form $\int dz^2 j^1(z) \bar{k}^3(\bar{z})$. The boundary conditions and asymptotic symmetries of supergravity on this background were further discussed in ref.~\cite{Compere:2013bya}.

Finally recall that the holographic duals to extremal black holes are expected to have the following two features, a geometry containing WAdS$_3$ in the bulk, and 
a $(2,1)$ deformation in the boundary. The fact that our construction of the $T\bar{K}$ deformation meets both expectations makes it a promising toy model of Kerr/CFT.
It also gives us a strong hint that the $T\bar{K}$ deformation may play an important role in understanding, or even defining, the nonlocal QFTs dual to (near) extremal black holes.


\section{String theory spectrum} \label{se:spectrum}

In this section we derive the finite-size spectrum of the $T\bar{K}$-deformed theory via spectral flow. In particular, we show that the single trace deformation reproduces the general form of the spectrum derived on the field theory side in~\cite{Guica:2017lia} up to a shift in the $\widebar{U(1)}_R$ charge.

\subsection{The undeformed spectrum on the cylinder} \label{se:spectrumcylinder}

To begin, let us put the dual CFT on the cylinder. This is accomplished by compactifying the boundary of the original background metric~\eqref{metric} such that 
  \eq{
  (\g,\, \bar{\g}) \sim (\g + 2 \pi R, \, \bar{\g} + 2\pi R). \label{compactification}
  }
In this case eq.~\eqref{metric} no longer describes the AdS$_3$ vacuum in Poincar\'e coordinates but rather the massless BTZ black hole. 

Now recall that string theory on an AdS$_3$ background features both discrete and continuous representations of $SL(2,R)$, as well as ``flowed'' representations obtained via spectral flow~\cite{Maldacena:2000hw}. The latter are parametrized by an additional integer $w$ which may be interpreted as the winding number of long strings corresponding to the continuous representation. Nevertheless, for either short or long strings, the flowed representations can be obtained by imposing the following boundary conditions on the worldsheet coordinates
  \eq{
  \g (\s + 2\pi) = \g (\s) + 2 \pi w R, \qquad \qquad   \bar{\g} (\s + 2\pi) = \bar{\g} (\s) +2 \pi w R. \label{winding}
  }
In the background~\eqref{metric} these boundary conditions lead to the following shift in the $L_n$ and $\bar{L}_n$ modes of the worldsheet stress tensor~\cite{Maldacena:2000hw}
  \eq{
  \L_n = L_n + w R \, j^{-}_n, \qquad \qquad \bar{\L}_n = \bar{L}_n - wR\, \bar{j}^-_n.
  }
That these shifts are proportional to the modes of the $j^{-}(z)$ and $\bar{j}^-(\bar{z})$ $SL(2,R)$ currents is to be expected, as the latter generate shifts in the $\g$ and $\bar{\g}$ coordinates. In particular, this means that the zero mode charges of $j^{-}(z)$ and $\bar{j}^-(\bar{z})$ correspond to the spacetime momentum $p$ and $\bar{p}$ conjugate to $\g$ and $\bar{\g}$, i.e.~we have
  \eq{
  j^-_0 = - \frac{1}{2\pi} \int dz \,j^{-}(z)= p, \qquad \qquad \bar{j}^-_0 = -\frac{1}{2\pi} \int d\bar{z} \, \bar{j}^{-} (\bar{z}) = \bar{p}.
  }
As a result of the spectral flow,  the Virasoro constraints receive a linear dependence on the momenta. The constraints on a state parametrized by the $SL(2,R)$ weight $j$ at level $N$ are thus given by
  \eq{
  0 &= \L_0 - 1 =  -\frac{j(j-1)}{k} + \dd_{S^3 \times {\cal M}^4} + N - 1 + w R p, \label{vircons1} \\
  0 &= \bar{\L}_0 - 1 =  -\frac{\bar{j}(\bar{j}-1)}{k} + \bar{\dd}_{S^3 \times {\cal M}^4} + \bar{N} - 1 - w R \bar{p}, \label{vircons2}
  }
where we assumed that $k \gg 1$, the first term in each equation is proportional to the Casimir of $SL(2,R)$, and $\dd_{S^3 \times {\cal M}^4}$, $\bar{\dd}_{S^3 \times {\cal M}^4}$ denote the contributions of $S^3 \times {\cal M}^4$.

As discussed in~\cite{Maldacena:2000hw}, the physical spectrum contains both discrete representations $\mathcal{D}^{\pm,w}_{j}$ with $1/2<j<(k-1)/2$ and principal continuous representations $\mathcal{C}_{j,\alpha}^w$ with $j=1/2+is$ and $s\in \mathbb{R}$. When $w<0$ (in our conventions) the left moving energy $E_L\equiv p$ is bounded from below for both the continuous and the discrete highest weight representation $\mathcal{D}^{+,w}_{j}$.\footnote{We thank Amit Giveon and Monica Guica for discussions on this point.}


\subsection{The deformed spectrum on the plane}

Once the $T\bar{K}$ deformation is turned on, the string theory spectrum receives corrections that depend on the dimensionless coupling constant $\l/R$. One way to derive the spectrum is to note that the deformed equations of motion and Virasoro constraints derived from the action~\eqref{deformedaction} are equivalent to the corresponding undeformed quantities derived from~\eqref{action} after a nonlocal field redefinition~\cite{Frolov:2005dj, Alday:2005ww, Azeyanagi:2012zd}. This corresponds to a nonlocal change of the spacetime coordinates given by
  \eq{
  \bp\hat{\g} &= \bp \g + \l \( \bp \psi + \cos\t\,\bp \vp \), \label{bpgamma} \\
   \p \hat{\g} &= \p \g - \l^2 e^{2\phi} \p \bar{\g}, \\
 \p \hat{\psi} &= \p\psi + 2 \l e^{2\phi} \p \bar{\g}, \\
 \bar{\p} \hat{\psi} &= \bar{\p} \psi. \label{bppsi}
 }
In terms of $\hat{\g}$ and $\hat{\psi}$ the deformed background fields~\eqref{deformedmetric} and~\eqref{deformedbfield} reduce to their undeformed, $\l = 0$ versions, but with modified boundary conditions,
  \eq{
  \hat{\g}(\s + 2\pi)  &= \hat{\g}(\s)  - 2\pi \frac{\l}{k} \( \bar{q} - \l p \), \label{newbc1}\\
   \hat{\bar\gamma}(\s + 2\pi) &= \hat{\bar \gamma}(\s), \\
  \hat{\psi}(\s + 2\pi) &= \hat{\psi}(\s) + 4 \pi \frac{\l}{k} p, \label{newbc2}
  }
where, following the conventions established in~\cite{Azeyanagi:2012zd}, $\bar{q}/2$ is the zero mode charge of the $\bar{k}^3(\bar{z})$ current\footnote{Note that up to normalization, the momenta $p$, $\bar{p}$, and the $k^3_0$ charge $\bar{q}$ are also the charges of the corresponding spacetime currents.}
  \eq{
  \bar{k}^3_0 \equiv \frac{i}{2\pi} \int d\bar{z} \, \bar{k}^3(\bar{z}) = \frac{\bar{q}}{2}.
  }

The twisted boundary conditions in eqs.~\eqref{newbc1} --~\eqref{newbc2} induce a shift in the left and right-moving $SL(2,R)$ and $SU(2)$ currents. This is equivalent to spectral flow of the current algebra. In particular, the $L_n$ and $\bar{L}_n$ modes of the worldsheet stress-energy tensor 
become
  \eq{
     \L_n & = L_n - \frac{\l}{k} \big (\bar{q} - \l p \big ) \, j^{-}_n, \\
   \bar{\L}_n & = \bar{L}_n + \frac{1}{k} \Big ( \l^2 p^2 \d_n - 2 \l p \, \bar{k}^3_n \Big ).
   }
   %
Thus, the Virasoro constraints of the deformed theory on the plane are given by 
  \eq{
  0 &= \L_0 - 1 =  -\frac{j(j-1)}{k} + \dd_{S^3 \times {\cal M}^4} + N - 1 - \frac{\l p}{k} \( \bar{q} - \l p \),  \label{deplane1} \\
  0 &= \bar{\L}_0 - 1 =  -\frac{\bar{j}(\bar{j}-1)}{k} + \bar{\dd}_{S^3 \times {\cal M}^4} + \bar{N} - 1  - \frac{\l p}{k} \( \bar{q} - \l p \). \label{deplane2}
  }
After the deformation, the theory remains invariant under the right-moving $\overline{SL(2,R)}_R$ symmetry. 
Thus, for highest weight representations of $\widebar{SL(2,R)}_R$ with $\bar{h} =\bar j$ the deformed conformal weight is given by
  \eq{
  {\bar h} (\l) &={\frac{1}{2}}+\sqrt{\({\bar h}(0)-{\frac{1}{2}}\)^2 - \lambda p \bar q + \lambda^2 p^2}. \label{conformalweight}
  }
  %
Note that the spacetime conformal weight also depends on the spacetime momentum $p$.
This is a generic feature of holography for NHEK and WAdS$_3$ spacetimes and a hint of the nonlocality of the dual field theory. For more details on the spectrum on the plane see ref.~\cite{Azeyanagi:2012zd}.


\subsection{The deformed spectrum on the cylinder}

In analogy to the discussion of the previous section, to obtain the spectrum of the deformed theory on the cylinder we need to perform an additional spectral flow. This amounts to an additional change of boundary conditions beyond that considered in eq.~\eqref{winding}. We thus arrive at the following twisted boundary conditions 
  \eq{
  \hat{\g}(\s + 2\pi)  &= \hat{\g}(\s) + 2\pi w R  - 2\pi \frac{\l}{k} \( \bar{q} - \l p \), \label{newbc3}\\
   \hat{\bar\gamma}(\s + 2\pi) &= \hat{\bar \gamma}(\s)+2\pi wR, \\
  \hat{\psi}(\s + 2\pi) &= \hat{\psi}(\s) + 4 \pi \frac{\l}{k} p. \label{newbc4}
  }
The aforementioned nonlocality in the change of spacetime coordinates is reflected in eqs.~\eqref{newbc3} and~\eqref{newbc4} in different ways. As noted above there is the dependence on the momentum $p$ and the charge $\bar{q}$, which are nonlocal in the coordinates. Furthermore, the beginning and end of the string do not differ by a multiple of $2\pi R$ for generic $\lambda$. 

The new boundary conditions are equivalent to spectral flow on both the left and right-moving $SL(2,R)$ sectors, as well as the right-moving $\widebar{SU(2)}_R$ sector. In particular, they lead to the following shifts in the modes of the worldsheet stress energy tensor and $\bar{k}^3(\bar{z})$ current
  \eq{
   \L_n & = L_n + \bigg [ w R - \frac{\l}{k} \big (\bar{q} - \l p \big ) \bigg ]\, j^{-}_n, \\
   \bar{\L}_n & = \bar{L}_n - w R\, \bar{j}^{-}_n + \frac{1}{k} \Big ( \l^2 p^2 \d_n - 2 \l p \, \bar{k}^3_n \Big ),\\
   {\bar{\K}}^{3}_n&={\bar k}^3_{n}-\lambda p\,\delta_{n,0},
  }
where $\bar{k}^3_n$ denotes the modes of the $\bar{k}^3(\bar{z})$ current. Thus, in the deformed theory on the cylinder the Virasoro constraints read
  \eq{
  0 &= \L_0 - 1 =  -\frac{j(j-1)}{k} + \dd_{S^3 \times {\cal M}^4} + N - 1 + w R p - \frac{\l p}{k} \( \bar{q} - \l p \),  \label{deformedvircons1} \\
  0 &= \bar{\L}_0 - 1 =  -\frac{\bar{j}(\bar{j}-1)}{k} + \bar{\dd}_{S^3 \times {\cal M}^4} + \bar{N} - 1 - w R \bar{p} - \frac{\l p}{k} \( \bar{q} - \l p \), \label{deformedvircons2}
  }
where we have used the fact that the zero mode charges of the $j^{-}(z)$ and $\bar{k}^3(\bar{z})$ currents are respectively given by $p$ and $\bar{q}/2$.

On the cylinder, the left and right moving energies $E_L(\l)$ and $E_R(\l)$, as well as the $\widebar{U(1)}_R$ charge $\bar{Q}(\l)$, are identified as the zero modes of the following currents,\footnote{As shown in~\cite{Azeyanagi:2012zd}, the spacetime charge associated with translations along $\psi$ is given by $\bar{q}$ which remains quantized after the deformation.}
  \eq{
   E_L(\lambda) & \equiv j^-_0 = p,\\
  E_R(\l) &\equiv -{\bar j}^-_0= -\bar{p}, \\
  \bar{Q}(\l) & \equiv 2 \bar{\K}^{3}_0   = {\bar{q}}{} - 2\l E_L(\l).\label{deformedcharge}
    }
We can now solve the Virasoro constraints before and after the deformation, and express the deformed energies in terms of the undeformed ones. In terms of these variables the spectrum of the $T\bar{K}$-deformed theory may be conveniently written as
  \eq{
  \!\!\big ( 1 - g_{w} \big ) E_L(\l) = E_L(0), \qquad E_R(\l) - g_{w}\, E_L(\l) = E_R(0), \label{deformedspectrum}
  }
where $g_w$ is proportional to the spectral flow induced by the deformation and is given by
  \eq{
   g_{w} = \frac{1}{k} \frac{\l}{w R} \Big [ \bar{q} - \l E_L(\l) \Big ].
  }
The solution to $E_L(\l)$ in eq.~\eqref{deformedspectrum} depends on the sign of the spectral flow parameter $w$ and reads
   \begin{subequations}
   \begin{empheq}[left={E_L(\l)=\empheqlbrace}]{align}
   & \dfrac{\l \bar{q} +k| w |R}{2 \l^2} - \dfrac{1}{2 \l^2} \sqrt{\vphantom{\tfrac{1}{2}} (\l \bar{q}+k |w| R)^2 - 4 k |w| R \,\l^2 E_L(0)\,},  &  w<0, \label{negativewinding} \\
& \dfrac{\l \bar{q} - k w R}{2 \l^2} + \dfrac{1}{2 \l^2} \sqrt{\vphantom{\tfrac{1}{2}} (\l \bar{q}- k w R)^2 + 4 k w R \,\l^2 E_L(0)\,}, &w>0.  \label{positivewinding}
  \end{empheq}
  \end{subequations}  %
As discussed towards the end of Section~\ref{se:spectrumcylinder} only the branch with $w<0$ is physical.   

In contrast to the left-moving energy $E_L(\l)$, the following combination of charges remain unchanged after the deformation
  \eq{
  2P(\l) \equiv E_L(\l) - E_R(\l) & = E_L(0) - E_R(0) = 2 P(0), \label{spec1} \\
  E_R(\l) + \frac{1}{4kwR} \bar{Q}(\l)^2 & = E_R(0) + \frac{1}{4k wR} \bar{Q}(0)^2, \label{spec3}
  }
where $P(\lambda)=P(0)$ denotes the momentum along the $\gamma^1$ direction.

To summarize, the $T\bar{K}$ deformation leads to a deformed spectrum characterized by the left-moving energy~\eqref{negativewinding}, the right-moving energy~\eqref{spec1}, and the $U(1)$ charge~\eqref{deformedcharge}. We conclude with the following observations on the spectrum:
\begin{enumerate}[leftmargin=20pt]
\item[i.] In our conventions, the field theory spectrum obtained in~\cite{Guica:2017lia} is given by\footnote{The dictionary is as follows: $\tilde{E}_R^{\mathrm{here}}\ra E_L^{\mathrm{there}}$, $\tilde{E}_L^{\mathrm{here}}\ra E_R^{\mathrm{there}}$.}
  \eq{
  \(1 - g \) \tilde{E}_L(\mu) = \tilde{E}_L(0),  \qquad \tilde{E}_R(\mu)-g \, \tilde{E}_L(\mu) = \tilde{E}_R(0),  \qquad g = \frac{\mu}{2R}\bar{q}, \label{spectrum}
  }
where $\mu$ is the deformation parameter. Eq.~\eqref{spectrum} takes the same form of eq.~\eqref{deformedspectrum}, with $\mu\ra{\frac{2}{wk}}\l$ and a shift in $\bar{q}$ given by $\bar{q} \ra \bar{q} - \l E_L(\l)$. 
\item[ii.]
The angular momentum $P(\l)$, which must take discrete values, is unchanged after the deformation. This agrees with the results from the spectrum derived in~\cite{Guica:2017lia}.
\item[iii.] For $\lambda=0$, both branches of $E_L(\l)$ in eqs.~\eqref{negativewinding} and~\eqref{positivewinding} satisfy the condition $E_L(\lambda \ra 0)=E_L(0)$.
\item[vi.] For $\lambda\neq0$, the physical branch with negative $w$~\eqref{negativewinding} becomes complex for large enough $E_L(0)$, while the branch with positive $w$ remains real.


\end{enumerate}



\section*{Acknowledgments}
We are grateful to Monica Guica for insightful discussions.  We also thank the Third ERC Solvay Workshop on ``Holography" where the project was initiated. This work was supported by the National Thousand-Young-Talents Program of China and NFSC Grant No.~11735001. The work of L. \!A. was also supported by the International Postdoc Program at Tsinghua University. 






\ifprstyle
	\bibliographystyle{apsrev4-1}
\else
	\bibliographystyle{JHEP}
\fi

\bibliography{jtbar}



\end{document}
